\documentclass[a4paper,twoside]{article}

\usepackage[top=1.7cm,bottom=1.2cm,left=2.8cm,right=2.3cm,
            includehead,includefoot]{geometry}
\usepackage{multicol}
\usepackage[perpage,symbol*]{footmisc}

\newcommand{\upcite}[1]{\textsuperscript{\cite{#1}}}  

\date{}

\begin{document}

\title{Progress on multi-waveband observations of supernova remnants\footnote{Supported by the National Natural Science Foundation of China (Grant No 10673018)}}

\author{Yang Xuejuan$^{1,2}$, Lu Fangjun$^{1}$, Tian Wenwu$^{3,4}$\footnote{To whom correspondence should be addressed. E-mail: tww@bao.ac.cn, tww@iras.ucalgary.ca}}

\maketitle
\begin{center}
1. Particle Astrophysics Center, Institute of High Energy Physics, CAS, Beijing 100049, P.R. China\\
2. Department of Astronomy, Beijing Normal University, Beijing 100875, P.R. China\\
3. National Astronomical Observatories, CAS, Beijing 100012, P.R. China\\
4. Department of Physics \& Astronomy, University of Calgary, Calgary, AB T2N 1N4, Canada\\
\end{center}
\begin{abstract}
A number of observational advances have increased our knowledge
about supernova remnants. In this paper we review the main
progresses made in the last decade, including new discoveries of
supernova remnants and the associated pulsars, nucleosynthesis, the
interaction between supernova remnants and molecular clouds, dust in
the supernova remnants, shock physics and cosmic ray accelerations.

{\bf keywords: Supernova remnant, observations}

\end{abstract}

Massive stars usually end their lives with supernova (SN)
explosions. It's very powerful, with total energy of $\sim 10^{44} J$,
to outshine a galaxy at peak. The outer layers of the exploding star
are ejected at supersonic speed, resulting in an outward blast wave.
Meanwhile, the blast wave is decelerated as expanding into the
interstellar medium (ISM), forming a reverse shock propagating
inwards. 
The shocked materials, along with the
stellar remnant (if exist) form a supernova remnant (SNR). SNRs are
often very bright in the radio and X-ray bands, as the shocks often
heat the ISM and ejecta to X-ray emitting temperature and accelerate
the electrons to produce synchrotron radiation in radio and/or even
X-ray bands. The interaction between an SNR and the nearby molecular
clouds (MCs) often trigger emission of molecular lines. The optical
emission of an SNR often comes from shocked ejecta, while infrared
emission traces the dust around. The polarization observations of SNRs
bring new insight to SNR physics\upcite{Xu07}.

 The SNe are classified as type I and II based on the presence of
Hydrogen Balmer lines in their spectra at maximum brightness or not,
and each type has sub-types for different properties of spectra or
lightcurves (c.f Fig. 1 in Ref\cite{Vink04a}). It's widely accepted
that SNe Ia correspond to the thermonuclear disruption of a C-O
dwarf in an accreting binary system after its mass appoaches
Chandrasecar limit. SN Ib/Ic and SNe II are from core collapses of
massive stars, giving birth to neutron stars for progenitor mass in
the range of $9M_{\odot} \sim 25M_{\odot}$ or black holes for more
massive stars\upcite{Heger03}. SNRs are important for understanding
our Galaxy. They heat up the interstellar medium, distribute heavy
elements throughout the Galaxy, and accelerate cosmic rays (CRs).
The shock wave of an SN injects energy into the interstellar gas,
compresses and accelerates it. The interaction between an SNR and
molecular clouds may trigger star formation. SNRs are also believed
to be the dominant source of Galactic CRs. 
In the last decade, great progresses have been made in understanding
SNRs, thanks to the new generation of telescopes.

\section{Supernova explosion and physics}
\subsection{Discoveries of SNRs and pulsars associated}

This discrepancy between the number of known Galactic SNRs ($\sim
270$) and that predicted by theory ( $>= 1000$ ) has been considered
as the result of the selection effects in the current
sensitivity-limited radio surveys. This is supported by the
discoveries of many new SNRs and candidates in the recent surveys
with high sensitivity and spatial resolution at low radio
frequencies\upcite{Helfand06, Brogan06, Kothes05}. Some individual
SNRs have newly been discovered from multi-band observations, e.g.
Tian et al.\upcite{Tian08b} by radio observations, Stupar et
al.\upcite{Stupar07} by optical observations, and Funk et
al.\upcite{Funk07} by X-ray observations.

In addition to the expanding SNRs, some core-collapse SN explosions
also produce pulsars. Discoveries of pulsars in SNRs are therefore
clues for the classification of the SNe associated. Up to now more
than 50 pulsars have been claimed to be likely associated with the
SNRs \upcite{Tian07b,Tian06}. New generation of X-ray space
observatories, e.g. XMM-Newton and Chandra X-ray observatories, have
shown their power in detecting  X-ray pulsars. With its sub-second
spatial resolution, Chandra has discovered pulsars in core-collapse
SNRs G292.0+1.8\upcite{Hughes03}, G54.1+0.3\upcite{Lu02},
G21.5-0.9\upcite{Camilo06} and pulsar wind nebula candidates in
N23\upcite{Hayato06}, G15.9+0.2\upcite{Reynolds06} and DA
530\upcite{Jiang07}.

One of the most exciting progresses about pulsars is the discoveries
of ``Anomalous X-ray Pulsars'' (AXPs) and ``Soft $\gamma$-ray
Repeaters'' (SGRs), which have very different properties from
traditional radio pulsars. AXPs and SGRs cannot be powered by the
rotational energy or by accretion of matter from a binary companion
star, and have extremely high surface magnetic fields ($B>10^{14}G$,
the ``magnetars'', see Ref\cite{Woods04} for a review). Some AXPs
and SGRs are associated with SNRs, e.g. Kes 73/AXP 1E 1841-045,
G29.6+0.1/AX J1845-0258, CTB 109/AXP 1E 2259+586, N49/SGR 0526-66
etc\upcite{Gaensler01,Gaensler03}. The compact object in the center of Cassiopiea A
(Cas A) might also be a magnetar\upcite{Krause05}. The newest exciting discovery is from 
the Fermi Gamma Ray Space Telescope which detects a $\gamma$-ray pulsation, i.e. first $\gamma$-ray-only pulsar with a period of 316.86 ms, near the center of SNR CTA 1\upcite{Abdo08}.

The association between SNRs and AXPs/SGRs could be used to
constrain the properties of AXPs/SGRs. Vink \& Kuiper\upcite{Vink06}
investigated the explosion energies of three SNRs (Kes 73, CTB 109
and N49) hosting AXPs/SGRs. They found that these SNRs' energies are
close to those of normal SNe and favor the possibility that
magnetars descend from progenitors with high magnetic field cores
instead of rapidly rotating proto-neutron stars\upcite{Vink06}.
However, this was argued against by the 50\% higher explosion
energy of Kes 73 \upcite{Tian08a}, which implies either a larger
magnetic field decay rate in the magnetar model or a larger accretion
rate in accretion-based models. The more high-resolution multi-waveband
observations of the AXPs/SGRs-related SNRs will help to constrain their
properties, and distinguish from different theoretical models.

\subsection{Nucleosynthesis}
Numerical calculations\upcite{Woosley95, Thielemann96} predict that
the nucleosynthesis during SN explosion is in onion type with
dominant elements ordered in shells following their atomic
number.

In young SNRs, the element stratification might be reserved since
the relatively short time of interaction with surroundings. A good
icon is the Tycho's SNR. X-ray observations show that Fe K line
peaks at a much smaller radius than those of Si, S and Fe
L\upcite{Hwang97,Decour01}. Hwang et al.\upcite{Hwang97} find that
the Fe K emission in Tycho is from an isolated component with
ionization age 100 times smaller than that of Si or S, implying that
Fe ejecta may retain some stratification and be located at the inner
layers thus reverse shocked more recently. X-ray spectroscopy of
G292.0+1.8 shows that it has little evidence of metal (Si, S and Fe)
enriched ejecta from explosive nucleosynthesis, suggesting that the
ejecta are strongly stratified by composition and that the reverse
shock has not propagated to the Si/S- or Fe- rich
zones\upcite{Park04}.

However, such stratification could be destroyed during the explosion
in some cases. For Cas A, the Fe-rich ejecta is located at larger
radius than that of Si\upcite{Hughes00, Willingale02}. It is
concluded that the ejecta has undergone a spatial inversion, which
might be caused by the neutrino-driven convection initiating
core-collapse\upcite{Hughes00}.

The numerical models have also calculated the nucleosynthesis
yield as a function of progenitor mass\upcite{Woosley95, Thielemann96}.
By comparing the abundance pattern from both observations and theoretical
calculations, the progenitor mass can be estimated\upcite{Ganzalez03}.
 However, the observational results basically come from the spectra
 fitting with plasma models, which only contain the relatively abundant
 elements that show strong emission lines in the spectra. As the increase of the sensitivity
 of the detectors, some new emission lines have been detected, such Cr, Mn
 and etc\upcite{Hwang00, Miceli06}, which are not included in all the available
 plasma models. Since these elements have been included in the numerical
 calculations, it will be helpful to constrain the properties of the SNe/SNRs
 including them in the plasma models.

\section{SNRs and their environment}
\subsection{Interaction with molecular clouds}
Core collapse SN explosions are expected to occur in MCs, since
their massive progenitors ($\geq 8 M_{\odot}$) are born in MCs and
their lifetimes ($\leq 3 \times 10^7$ years) are often shorter than
the typical lifetime of an MC\upcite{Chevalier01}. As the SNRs
expand, they might interact with MCs.

The first clear evidence for this interaction is from IC443 based on
observations of shocked CO\upcite{DeNoyer79a} and
OH\upcite{DeNoyer79b} emission. Such study became more intense after
the realization of OH 1720 MHz maser line emission as a ``signpost''
of the interaction\upcite{Frail94} ( see Ref\cite{Wardle02} for a
review). Surveys have been done \cite{Frail96,Green97} and \cite{Koralesky98} and several SNRs have been found with such maser emission, including W28, W44,
3C391 etc. These SNRs are mostly mixed-morphology SNRs, which have been suggested to be strongly associated with the OH 1720 MHz maser emission \cite{Yusef03} and \cite{Chen04}. It is predicted that the OH 1720 MHz line will switch off at large OH column density ($N_{OH}$), and the 6049 MHz and 4765 MHz lines will be on instead, with a peak $N_{OH}$ of 3 x $10^{17} cm^{-2}$,  and of several times higher, respectively \cite{Wardle07} and \cite{Pihl08}. These lines may serve as a
complementary signal of warm, shocked gas when the OH column density
is large.

Such interaction can also be traced by the shocked emission lines
from CO, H$_2$ or other molecules. Using these diagnosis,
interactions are identified in G347.3-0.5\upcite{Moriguch05} and
HB21\upcite{Byun06} etc.

The direct way to identify the
interaction is to determine the distances to an SNR and MC system.
Based on a new distance-measurement method (Tian-Leahy method by the HI and
CO observations), Tian et al.\upcite{Tian07c} suggest the interaction between SNR
G18.8+0.3 and a molecular cloud, and give a distance of $\sim$ 12 kpc to the SNR/CO
cloud system. A more reliable example is the SNR W41/HESS 1834-087/molecular cloud system.
High-precision distance measurements to the system support that the SNR is physically
associated with the giant molecular clouds so the SNR/cloud interaction leads to the TeV
$\gamma$-ray emission in the cloud material\upcite{Tian07d}. The method is so powerful
that an intriguing puzzle on the distance to SNR Kes 75/PWN J1846-0258 system has been solved
recently\upcite{Leahy08}. It is worth determining distances to more claimed SNR/could
systems by this method so that we could refine current models of SNR/cloud interaction.

\subsection{Dust in SNRs}
It has been discovered that there is a huge amount of dust ($10^8
\sim 10^9 M_{\odot}$) in very high redshifted ($z>6$) galaxies and
quasars\upcite{Isaak02, Bertoldi03}, corresponding to the Universe
age of 700 million years. The stellar winds at the late time
evolution of the stars are thought to be the main sources of dust in
galaxies, but they are not able to produce that much in such a short
time\upcite{Dwek98}. Type II SNe are potential sources, with a dust
production of $0.08 \sim 1 M_{\odot}$ in the ejecta per SN, varying
with metallicity and progenitor mass\upcite{Todini01}.

Dunne et al.\upcite{Dunne03} report a detection of cold dust of
$\sim 2-4M_{\odot}$ in Cas A. They imply that SNe are at least as
important as stellar wind in producing dust in our Galaxy and would
have been the dominate source of dust at high redshift\upcite{Dunne03}.
The optical and mid-infrared observations of SN 2003gd show a total dust amount
of 0.02$M_{\odot}$, suggesting that SNe might be major dust
factories\upcite{Sugerman06}. However, it is then argued that the
dust detected in Cas A may originate from interstellar dust in a
molecular cloud complex located in the line of sight between the
Earth and the SNR\upcite{Krause04}. The Spitzer observations show
that the dust mass in SN 2003gd is only about $4 \times 10^{-5}
M_{\odot}$, arguing against the presence of 0.02$M_{\odot}$ of newly
formed dust in the ejecta\upcite{Meikle07}. Recently Rho et
al.\upcite{Rho07} present a comprehensive analysis of the dust mass
in Cas A with the Spitzer observations and show that the total dust
mass is sufficient to explain the lower limit of the dust mass in
high redshift galaxies. It still remains an open question what's the
real total amount of dust in the ejecta of core collapse SNe, and
more investigations are required.

\section{Shock physics}
\subsection{Electron-Ion temperature equilibrium}
The shocks in SNRs are often referred to as collisionless shocks as
the particle collision length scale is much larger than the typical
size of the shock structure.
Although the nature of
electron and ion heating behind collisionless shocks in SNRs remains
an open question, a number of observational advances have increased
our knowledge about it.

Behind the collisionless shock there can exist a population of cold
neutral ions that are not affected by the shock passage. Some of
them might be collisionally excited before being destroyed by
collisional ionization or charge transfer.
They will emit narrow H$\alpha$ and H$\beta$ Balmer lines with widths
representing the pre-shock temperatures. Those having charge
exchange with shock heated protons will produce broad Balmer lines,
whose widths will be given by the temperature of postshock
protons.
In this case, the broad-to-narrow line ratio can be used to measure the
electron-proton temperature ratio ($T_e/T_p$), which represents the
equilibration degree. Ghavamian et al.\upcite{Ghavamian01} present
observations and theoretical studies of the Cygnus Loop, RCW 86 and
Tycho based on this method, and get the equilibration degree of
electron/proton temperatures. However, the narrow Balmer lines might
be contaminated by emission from a shock precursor, leading to a
higher modeled broad-to-narrow ratio, such as for DEM
L71\upcite{Ghavamian03}. In such case, the electron temperature
can be determined from the X-ray bremsstrahlung emission
instead\upcite{Ghavamian03, Rakowski03}.

Using the high spectral resolution data from the XMM-Newton
observations, Vink et al.\upcite{Vink03a} find a 3.4 eV broadening
of O emission line in SN 1006, implying an O temperature of
$530\pm150$ keV. The electron temperature from the same observation
is about 1.5 keV, suggesting a low degree of equilibration in this
remnant. This is consistent with the optical\upcite{Ghavamian02} and
ultraviolet\upcite{Laming96} studies. Rakowski\upcite{Rakowski05}
compile each method that has been used to measure the equilibration
and every SNR on which they have been tested. A negative correlation
between the degree of equilibration and the shock velocity is
indicated\upcite{Rakowski05}. Such correlation is further studied
by Ghavamian et al.\upcite{Ghavamian07}. They find a relationship
$(T_e/T_p)_0\propto v_s^{-2}$ that is fully consistent with the observations,
where $(T_e/T_p)_0$ represents the electron-proton temperature equilibration
at the shock front and $v_s$ the shock velocity. Although such a
relationship has been suggested, many questions are remained: What
mechanism causes the sharp decrease of temperature equilibrium at
small shock speed? Does this relationship hold for collisionless shocks
in fully ionized gas? Progress in this developing field depends on accurate
modeling of the emission from pre-shock and post-shock gas, as well as
evidence from multiple wave-bands, and useful assessments of the cosmic ray
production and its effect on the shock. 

\subsection{Cosmic-Ray acceleration}
SNRs are believed to be the dominant source of Galactic CRs, at
least for energies up to the ``knee'' of the CR spectrum ($3 \times
10^{15}$ eV). The radio synchrotron emission at the shocks of
shell-type SNRs has provided direct evidence for accelerated
electrons with energies up to GeV range. It was raised to 10 $\sim$
100 TeV after the first detection of X-ray synchrotron filament in
SN1006\upcite{Koyama95}. Such filaments have also been detected in
Cas A, RCW 86, Tycho, Kepler, G266.2-1.2, G347.3-0.5 etc (see also
Ref\cite{Vink05}).

As the development of atmospheric Cerenkov detectors (H.E.S.S,
CANGROO series etc), it is able to image SNRs by TeV $\gamma$-ray
observations, which play an important role to trace the CR
acceleration in SNRs. $\gamma$-ray emissions are detected in several
SNRs, including RX J1713.7-3946, RX J0852.0-4622, G0.9+0.1, W41 etc
(see Ref\cite{Rowell06} for a H.E.S.S observation review). However,
the H.E.S.S observations to SN1006 have no detection of TeV
$\gamma$-ray emission from any compact or extended region associated
with the remnant\upcite{Aharonian05b}. Using the observed X-ray flux
and $\gamma$-ray upper limit, they get a lower limit on the
post-shock magnetic field of $B > 25\mu$G\upcite{Aharonian05b}.

 Various models have been established to account for the overall
spectra of nonthermal emission (from radio up to TeV $\gamma$-ray)
in SNRs, which can basically categorized as time-dependent and steady
ones. Fang et al.\upcite{Fang08} and Zhang et al.\upcite{Zhang07} model the
non-thermal emission from old and young supernova remnants under
time-dependent frame respectively. These models are applied to the
observations of SNRs, and can well represent their multi-wavelength
spectra. 

Although there is a lot of observational evidences for electron
acceleration in SNRs, that for proton (dominate component of CRs)
acceleration is rare. One example is RX J1713.7-3946, where pions
($\pi^0$) decay (the signature of proton acceleration) was
detected\upcite{Enomoto02}.

The amplification of magnetic field is potentially the key for
accelerating protons and heavier ions up to the ``knee'' of the CR
spectrum\upcite{Ballet06, Uchiyama07}. The X-ray synchrotron
filaments provide not only evidence of CR acceleration, but also
information of the magnetic field at the shock front. Based on the
width of the filaments, Vink et al.\upcite{Vink03b} estimated the
magnetic field strength to be $0.08 \sim 0.16 mG$ at the shock front
of Cas A, much higher than the Galactic average value ($\sim
3\mu$G). X-ray filaments in SN1006\upcite{Bamba03},
Tycho\upcite{Hwang02} and Kepler\upcite{Cassam04} indicate similar
magnetic field strength in these remnants, which might be evidence
for CR induced magnetic field amplification\upcite{Vink05}. Uchiyama
et al.\upcite{Uchiyama07} report the discovery of the brightening
and decay of X-ray hot spots in RX J1713.7-3946 on a one-year
timescale, which might imply that we have witnessed the ongoing
shock-acceleration of electrons in real time. They conclude that the
rapid variability shows the origin of X-rays to be synchrotron
emission of ultrarelativistic electrons, meanwhile the electron
acceleration occurs in a strong magnetic field with an amplification
factor of more than 100\upcite{Uchiyama07}.

\section{Summary and prospectives}
In this paper, we review the progresses in the study of SNRs from
multi-band observations. More SNRs and (rare type of) pulsars
associated are discovered. Using the high spatial and energy
resolution data of SNRs,  we are able to study the nucleosynthesis
process during stellar evolution and SN explosion, and also the
interaction between SNRs and their surroundings. As a perfect
laboratory of shock physics, we know more about the collisionless
shock in SNRs, such as the electron-ion equilibration after shock
passage and the cosmic-ray acceleration.

Nevertheless, there are still many open questions. The known basic
parameters of many SNRs, such as the distances and ages, have large
uncertainty. High-precision measurement to these parameters, however,
are very important to study not only the SNRs themselves,
but also the related pulsars, molecular clouds and so on. This depends on
further advances in both observational techniques and measurement methods. 
Although it's widely believed that SN Ia is from the thermonuclear explosion of a
white dwarf, is the explosion mechanism accretion single degenerate
or double degenerate? Is the explosion a detonation (supersonic),
deflagration (subsonic), or is there transition from deflagration to
detonation? By comparing the abundance pattern of SNRs with numerical
calculations, some information of the SN explosions has been given,
such as the explosion type, progenitor mass. However, the
current theoretical calculations are generally simplified with
respect to the real SN explosions, and thus often cannot match all
the observations. Meanwhile, as the discovery of emission lines from
less abundant elements, such as Cr or Mn, plasma models
which includes these elements would be required.  AXPs/SGRs have been
suggested to be neutron stars with very strong magnetic fields (``magnetars'').
Do they and high magnetic field radio pulsars different phases of a more
uniform class of object? What is their evolutionary sequence? Why are the spin
periods of AXPs/SGRs strongly clustered at 5$-$12 s? To answer these questions,
it would be helpful to enlarge the sample of this rare type of pulsars and make
high-precision observations. 
SNR shocks can accelerate CRs. But how efficient the acceleration is?
What is the maximum energy shocks can accelerated to? How strong is the
magnetic field? Is it amplified or just compressed? If the
amplification exists, what's the mechanism? The new generation of
telescopes will shed new light on these issues.

\thanks{We acknowledge support from the Natural Science Foundation of China.
We thank Drs. Han Jinling and Chen Yang for their informative comments on the paper.}

\end{document}